\documentclass[12pt]{article}
\usepackage[top=25truemm,bottom=27truemm,left=22truemm,right=22truemm]{geometry}

\usepackage{indentfirst}
\usepackage[dvipdfmx]{graphicx}
\usepackage{float}
\usepackage{url}
\usepackage{siunitx}
\usepackage{array}
\usepackage{amsmath}
\usepackage{amssymb}
\usepackage{fancybox}
\usepackage{ascmac}
\usepackage{bm}
\usepackage{mathrsfs}
\usepackage{cancel}
\usepackage{physics}
\usepackage{color}
\usepackage{authblk}
\usepackage{appendix}
\usepackage{comment}
\usepackage{mhchem}
\usepackage{caption,latexsym,amsfonts,mathtools,here}
\usepackage{subcaption}
\usepackage{hyperref}
\usepackage{multirow}

\numberwithin{equation}{section}

\begin{document}
\begin{titlepage}
\unitlength = 1mm
\begin{flushright}
YITP-25-105, KOBE-COSMO-25-11
\end{flushright}

\vskip 1cm
\begin{center}

{ \large \textbf{Enhancing photon-axion conversion probability\\ with squeezed coherent states
}}
\vspace{1.8cm}
\\
Taiki Ikeda$^*$, Sugumi Kanno$^{*\natural}$, and  Jiro Soda$^{\flat}$
\vspace{1cm}

%\shortstack[l]
{\it $^*$ Department of Physics, Kyushu University, Fukuoka 819-0395, Japan \\
$^\natural$ Center for Gravitational Physics and Quantum Information, Yukawa Institute for Theoretical Physics, Kyoto University, Kitashirakawa Oiwakecho, Sakyo-ku, Kyoto 606-8502, Japan \\ 
\it $^\flat$ Department of Physics, Kobe University, Kobe 657-8501, Japan
}

\vskip 4.0cm

{\large Abstract}\\
\end{center}

In particle physics,  axions and axion-like particles are ubiquitous.
Remarkably, ultra-light axions could constitute dark matter or dark energy.
Therefore, it is important to detect axions experimentally.
In the presence of a magnetic field, a photon can be converted into an axion,
and vice versa. Utilizing the conversion phenomenon, several methods for detecting axions
have been proposed. To improve detectability, it is desirable to use quantum sensing. However, since the conversion process is usually treated as classical wave dynamics,
it is unclear how to incorporate quantum effects such as entanglement. In this work, we formulate the photon-axion conversion in a quantum field theoretical manner. As a result, we succeed in evaluating the conversion probability from a photon quantum state to an axion quantum state. In particular, it turns out that squeezed coherent states can enhance the conversion probability.

\vspace{1.0cm}
\end{titlepage}

\hrule height 0.075mm depth 0.075mm width 165mm
\tableofcontents
\vspace{1.0cm}
\hrule height 0.075mm depth 0.075mm width 165mm

\section{Introduction}

Axions were originally introduced to solve the strong CP problem in quantum chromodynamics 
\cite{Peccei:1977hh, Weinberg:1977ma, Wilczek:1977pj, Kim:1979if, Shifman:1979if, Dine:1981rt, Zhitnitsky:1980tq}. 
Interestingly, pseudoscalar particles also appear ubiquitously in string theory~\cite{Svrcek:2006yi}, and we refer to these pseudoscalar particles simply as axions. The axions play an important role in cosmology, as they are a strong candidate for dark matter~\cite{Marsh:2015xka}. In particular, heavy axions can naturally drive slow-roll inflation  \cite{Freese:1990rb} due to their shift symmetry, while lighter axions are excellent candidate for dark matter~\cite{Preskill:1982cy, Abbott:1982af, Dine:1982ah, Hui:2016ltb}.
Remarkably, axions with an extremely small mass around $10^{-33}~{\rm eV}$ can even mimic the effects of a  cosmological constant \cite{Frieman:1995pm}. 
Given their broad implications across cosmology, detecting the axions experimentally is of great importance.

The axion field $\phi$, with a mass $m$, interacts with the electromagnetic fields through a Chern-Simons coupling of the form
$
     g\,\phi\,{\bf E}\cdot {\bf B} 
$, 
where $g$, ${\bf E}$ and  ${\bf B}$ are axion-photon coupling constant, the electric and magnetic fields, respectively. 
The coupling constant $g$ is experimentally constrained to be less than $10^{-10}\,{\rm GeV}^{-1}$.
Intriguingly, in the presence of a magnetic field, this interaction enables the conversion of photons into axions and vice versa~\cite{Raffelt:1987im}. 
The probability for a photon $\gamma$, with energy $E_\gamma$ to convert into an axion is given by the expression
\begin{eqnarray}
   P(\gamma \rightarrow \phi ) 
   = (\Delta_M L)^2\,\frac{\sin^2 \frac{\Delta_{\rm osc}}{2}L}{\left(\frac{\Delta_{\rm osc}}{2} L \right)^2} \ ,
   \label{probability1}
\end{eqnarray}
where $\Delta_M$ is a parameter that quantifies the mixing strength between photons and axions in an external magnetic field, $L$ is the propagation distance of the photon and $\Delta_{\rm osc}$ is the effective oscillation wavenumber which determins the oscillation distance of the photon-axion oscillation. The $\Delta_M$ and $\Delta_{\rm osc}$ are defined as 
\begin{eqnarray}
    \Delta_M  =\frac{1}{2} g B_T
      \ , \qquad
    \Delta_{\rm osc} =\sqrt{\left(\frac{m^2}{2 E_\gamma}\right)^2 + 4\Delta_M^2} \ ,
    \label{Deltas}
\end{eqnarray}
where $B_T$ is the component of the external magnetic field transverse to the photon's direction of motion. In laboratory experiments, the dimensionless quantity $gB_TL$ is typically much smaller than  $m^2 L/2E_\gamma$. Plugging (\ref{Deltas}) into (\ref{probability1}), the argument of the sine function is
\begin{eqnarray}
\frac{\Delta_{\rm osc}}{2}L=\frac{1}{2}\sqrt{\left(\frac{m^2L}{2E_\gamma}\right)^2+\left(gB_TL\right)^2}\,.
\end{eqnarray}
In the case of typical QCD axions, we can estimate as follows
\begin{eqnarray}
  \frac{m^2 L}{2E_\gamma}  
  =10^{4} \left(\frac{m}{10^{-6}\,{\rm eV}}\right)^2 
  \left(\frac{E_\gamma}{10^{-6}\,{\rm eV}} \right)^{-1} 
  \left(\frac{L}{1\,{\rm km}}\right)
\end{eqnarray}
and 
\begin{eqnarray}
  gB_TL  
  =10^{-6} \left(\frac{g}{10^{-10}\,{\rm GeV}^{-1}}\right) 
  \left(\frac{B_T}{10\,{\rm T}} \right) 
  \left(\frac{L}{1\,{\rm km}}\right) \ .
\end{eqnarray}
Thus, the conversion probability can be found as
\begin{eqnarray}
   P(\gamma \rightarrow \phi ) 
   = (\Delta_M L)^2 \frac{\sin^2 \frac{m^2}{4E_\gamma }L}{\left(\frac{m^2}{4E_\gamma } L \right)^2} \ .
\end{eqnarray}
The photon-axion conversion phenomenon~\cite{Sikivie:1983ip, Irastorza:2018dyq} has been employed in experimental searches for both solar axions \cite{CAST:2017uph} and axion dark matter \cite{ADMX:2009iij}. Photon-axion conversion has been widely discussed in both cosmological and astrophysical contexts~\cite{Mirizzi:2009aj, Masaki:2017aea,Nomura:2022zyy}.

Photon-axion conversion is typically analyzed within the framework of classical wave equations. Hence, the nonclassical nature of photons cannot be taken into account.
In the context of quantum sensing, nonclassical states of light are often used to enhance detector sensitivity~\cite{RevModPhys.89.035002}. 
The purpose of this paper is to reconsider the photon-axion conversion process from a quantum field theoretical perspective. Based on previous works ~\cite{Blasone:1996pn,Binger:1999nj,Blasone:2001du},
axion-photon mixing is analyzed from a quantum field theoretical perspective in ~\cite{Capolupo:2019xyd}. However, that analysis relies on the result of \cite{Raffelt:1987im}. In this paper, we consider the simplest case and formulate the problem from first principles. This formalism allows us to 
take into account nonclassical photons. 
Thus, we can determine whether nonclassical states can enhance the conversion probability. It turns out that
the photon-axion conversion probability
is enhanced for the squeezed coherent states.

The organization of the paper is as follows. In section 2, we introduce the photon-axion model and show that the action can be reduced into a coupled two scalar porblem. In section 3, we reproduce the conventional  conversion probability in a quantum field theoretical manner. In section 4, we apply the formalism to the case of nonclassical photons such as the squeezed state. Remarkably, we show the enhancement of the conversion probability for the squeezed coherent state. 
The final section  is devoted to the conclusion.

\section{Photon-axion system  }

Since axions interact weakly with other particles, it is challenging to detect them experimentally. To address this, photons are useful because they can be controlled with high precision and can be prepared in various quantum states such as coherent and squeezed states.  

We consider the action for electromagnetic fields coupled to  an axion field via a Chern-Simons term:
\begin{eqnarray}
S
=\int d^4x \,
\left[
-\frac{1}{2} \partial_\mu \phi \ \partial^\mu \phi -\frac{1}{2} m^2 \phi^2 
-\frac{1}{4}
 F_{\mu\nu}   F^{\mu\nu}
-\frac{g}{8}
\phi \ \epsilon^{\mu\nu\lambda\rho}  F_{\mu\nu}
F_{\lambda\rho}
\right]
\label{original action}\,,
\end{eqnarray}
where $m $ is the mass of axion, $g$ is axion-photon coupling constant, and $\epsilon^{\mu\nu\lambda\rho}$ is a Levi-Civita tensor with a convention $\epsilon^{0123}=-1$. 
The gauge field $A_\mu$ represents photons and the field strength is defined by $F_{\mu\nu}=\partial_\mu A_{\nu}-\partial_\nu A_{\mu}$. Since we consider photons with energies much less than 1 MeV, where electron-positoron pair creation does not occur, we ignore the Euler-Heisenberg term arising from the electron one-loop effect.
We also assume that plasma effects are negligible.

The equations of motion are derived from Eq.~(\ref{original action}) such as
\begin{eqnarray}
&&   \partial_\mu \partial ^\mu \phi -m^2 \phi =
   \frac{g}{8}
 \epsilon^{\mu\nu\lambda\rho}  F_{\mu\nu}
F_{\lambda\rho}  \label{eom:phi}\ ,\\
&&  \partial_\mu F^{\mu\nu} =
-\frac{g}{2}
\partial_\mu  \phi \ \epsilon^{\mu\nu\lambda\rho}  
F_{\lambda\rho}  \label{eom:F}\ .
\end{eqnarray}
We consider the situation in which a constant external magnetic field is applied in the laboratory. Then we can verify that the following are solutions:
\begin{eqnarray}
\phi =0 \ , \quad  {\bf E} =0 , \quad {\bf B} = {\rm const}.
\end{eqnarray}

In the following, we consider the quantum evolution of photons and axions in the above background field.

\subsection{Free axions}

The free part of the axion field action is given by
\begin{eqnarray}
S_\phi=\frac{1}{2}\int d^4x\ \left[\dot{\phi}^{ 2}-(\partial_i \phi)^2
-m^2 \phi^2 \right]\,,
\label{action:phi}
\end{eqnarray}
where the dot and $\partial_i$ denote the time and spatial derivatives, respectively.

\subsection{Free photons}

Next, we consider the action for the photon field, expanded to second order in perturbations $A_i$, which is given by
\begin{eqnarray}
S_A=\frac{1}{2}\int d^4x\  \left[\dot{A}_i^{2}
- (\partial_k A_{i})^2 \right]\,,
\label{action:A}
\end{eqnarray}
where the photon field satisfies the Coulomb gauge conditions $A^0 =0$ and $A^i{}_{,i}=0$.

We note that the electric field is represented by
\begin{eqnarray}
    {\bf E}  = - \nabla A^0 - \dot{\bf A} \,.\label{E}
\end{eqnarray}
%Then varying the actions (\ref{action:phi}) and (\ref{action:A}) with respect to $\phi$ and $A_i$, respectively, yields the following linear equations of motion in vector form:
Eqs.~(\ref{eom:phi}) and (\ref{eom:F}) give
\begin{eqnarray}
&&  -\ddot{\phi} + \partial_i^2 \phi -m^2\phi= g {\bf E}\cdot {\bf B} \ , \\
&& \nabla \cdot {\bf E} =  g \nabla \phi \cdot {\bf B}  \label{divergenceE}
\ ,\\
&&  \frac{\partial {\bf E}}{\partial t}
= \nabla \times {\bf B} +g \frac{\partial \phi}{\partial t}{\bf B}+g\nabla\phi\times{\bf E}
\ .
\end{eqnarray}
As seen from Eq.~(\ref{divergenceE}), we cannot impose the Coulomb gauge $A^0=0$ unless $\nabla \phi \cdot {\bf B}=0$.   
Therefore, in our setup, we consider an external magnetic field that is transverse to the direction of photon and axion propagation, such that ${\bm k} \cdot {\bf B}=0$, where ${\bm k}$ is the wave number vector of the photons and axions. 
Alternatively, one can consider the action for $A^0$ separately since
the equation of motion for the propagation of $A^i$ is decoupled from $A^0$. 
Therefore, in the following, the electric field (\ref{E}) is written by ${\bf E}=-\dot{\bf A}$.

\subsection{Photon-axion interaction}
The action for the photon-axion interaction, expanded to  second order in perturbations $\phi$ and $A_i$, is given by
\begin{eqnarray}
\delta S_{\rm I}= g B_i  \int d^4 x\,
   \phi\,\dot{A}_{i}
   = g B \int d^4x \ 
  \  \phi \ \dot{\psi}     \,,
\label{action:I}
\end{eqnarray}
where $B_i=\varepsilon_{ij\ell}\,\partial_j A_\ell $ is the  constant background magnetic field and $B\equiv|{\bf B}|$. We defined the electric field projected along the magnetic field as
\begin{eqnarray}
    \psi (t , x^i)= \frac{B^i}{B} A^i 
    \label{psi-A}\ .
\end{eqnarray}
Due to the coupling between the axion and electric fields, the interaction induces photon polarization.
Note that the action for $\psi$ is given by
\begin{eqnarray}
S_\psi=\frac{1}{2}\int d^4x\  \left[\dot{\psi}^{2}
- (\partial_i \psi)^2 \right]\,.
\label{action:psi}
\end{eqnarray}

\subsection{Action for the total system}

The quadratic action
in the background of the constant magnetic field reads
\begin{eqnarray}
  S =
  \frac{1}{2}\int dt d^3x\  \left[
  \dot{\phi}^{ 2}-(\partial_i \phi)^2
-m^2 \phi^2
 + \dot{\psi}^{2}
- (\partial_i \psi)^2 
+gB \phi \ \dot{\psi} -gB \psi \dot{\phi} \right]\,,  
\label{total-action}
\end{eqnarray}
where we have symmetrized the interaction terms with respect to $\phi$ and $\psi$.
%Then the canonical momenta of $\phi$ and $\psi$ derived from the above action are given by
%\begin{eqnarray}
%    \pi_\phi = \dot{\phi} -\frac{1}{2}gB \psi  \ , \quad
%    \pi_\psi = \dot{\psi} +\frac{1}{2}gB \phi \ .
%\end{eqnarray}
%The Hamiltonian can be calculated as 
%\begin{eqnarray}
%    H = \frac{1}{2}\int  d^3x\  \left[
%  \left( \pi_\phi +\frac{1}{2} gB \psi \right)^{ 2}
%  +(\partial_i \phi)^2
%+m^2 \phi^2
% + \left( \pi_\psi - \frac{1}{2} gB \phi \right)^{2}
%+ (\partial_i \psi)^2  \right]\,. 
%\end{eqnarray}
%Using this Hamiltonian, the system can be quantized. However, 
%in the next section, we take another route using the fact that $gB$ is smallvcompared to the inverse of a typical length scale.

\section{Coversion of a single photon into an axion}

In this section, we try to reproduce the conventional conversion probability from the perspective of quantum field theory. We also obtain higher-order corrections.

\subsection{Interaction Hamiltonian}

We adopt the interaction picture.
The free part of the action in Eq.~(\ref{total-action}) is given by
\begin{eqnarray}
  S =
  \frac{1}{2}\int dt d^3x\  \left[
  \dot{\phi}^{ 2}-(\partial_i \phi)^2
-m^2 \phi^2
 + \dot{\psi}^{2}
- (\partial_i \psi)^2 
\right]\,. 
\end{eqnarray}
Using annihilation and creation operators $a_{\bm k}$ and  $a^\dagger_{\bm k}$, we can expand the axion field operator as follows 
\begin{align}
\phi (t , x^i)= \frac{1}{(2\pi)^{3/2}}
\int d^3 k\, \left[ \phi_{\bm k}(t) a_{\bm k} \, 
+ \phi_{\bm k}^* (t) a_{-\bm k}^\dagger \right]e^{i\bf{k}\cdot\bm{x}}
\label{operator_phi}
\ ,
\end{align}
where we defined the mode function as
\begin{eqnarray}
    \phi_{\bm k}(t)  
    = \frac{1}{\sqrt{2\omega_\phi}} \ e^{-i\omega_\phi t}  \ ,
\end{eqnarray}
with the dipersion relation given by $\omega_\phi^2 = {\bm k}^2 +m^2 $.

Similarly for the electric field, using the mode function defined by
\begin{eqnarray}
    \psi_{\bm k}(t)  
    = \frac{1}{\sqrt{2\omega_\psi}} e^{-i\omega_\psi t}  \ ,
\end{eqnarray}
the electric field is expanded by using annihilation and creation operators $b_{\bm k}$ and  $b^\dagger_{\bm k}$
\begin{align}
\psi (t , x^i)= \frac{1}{(2\pi)^{3/2}}
\int d^3 k\, \left[ \psi_{\bm k}(t) b_{\bm k} \, 
+ \psi_{\bm k}^* (t) b_{-\bm k}^\dagger \right]e^{i\bf{k}\cdot\bm{x}}
\label{operator_psi}
\ ,
\end{align}
with the dispersion relation given by $\omega_\psi = k =|{\bm k}|$.

The interaction Hamiltonian derived from Eq.~(\ref{total-action}) is given by
\begin{eqnarray}
  H_I =
  -\frac{1}{2}gB\int  d^3x\  \left[
 \phi \ \dot{\psi} - \psi \dot{\phi} \right]\, . 
 \label{IH}
\end{eqnarray}
This interaction generates mixing between the photon and the  axion. It should be noted that axion-photon oscillation, as a  quantum mixing phenomenon, is fundamentally different from neutrino oscillation: axion-photon mixing is driven by a coupling in the presence of an external magnetic field,  whereas neutrino oscillation arises from the mixing between  mass and flavor eigenstates.

The time evolution operator for the system is calculated from the interaction Hamiltonian in the interaction picture as follows.
%\begin{eqnarray}
%U(t,0) &=& {\bf T} \exp \left[ -i\int^t_0 H_I (t) dt\right] \nonumber\\
%&=& 1-i \int^t_0 H_I (t_1 ) dt_1 + (-i)^2 \int^t_{0} H_I (t_2)  dt_2\int^{t_2}_0 H_I (t_1) dt_1 \nonumber\\
%&&  \qquad+ (-i)^3 \int^t_0 H_I (t_3) dt_3 \int^{t_3}_0 H_I (t_2) dt_2 \int^{t_2}_0 H_I (t_1) dt_1  + \cdots  
%\end{eqnarray}
%where ${\bf T}$ denotes the time ordered product. 
%In the present case, since the time integral factorizes, we can simplify it as
\begin{eqnarray}
U(t,0) = \exp \left[ -i\int^t_0 H_I (t) dt\right] 
= \exp[ -iQ]\,.
\end{eqnarray}
Substituting the operators (\ref{operator_phi}) and (\ref{operator_psi}) into the interaction Hamiltonian (\ref{IH}), we obtain
\begin{eqnarray}
 && Q = \int^t_0 H_I dt \nonumber\\
 &&\quad =
  -\frac{i}{2}gB\int  d^3k\ \left[
  U_{\bm k} f(t)\,
 b_{\bm k}^{\dagger} a_{\bm k} -   U_{\bm k}f^* (t)\,
 a_{\bm k}^{\dagger} b_{\bm k} + V_{\bm k} g(t)\, 
 b_{\bm k} a_{-\bm k} - V_{\bm k}g^*(t)\,
 a_{\bm k}^{\dagger} b_{-\bm k}^{\dagger}\right]\, ,
 \qquad
\end{eqnarray}
where we defined parameters
\begin{eqnarray}
U_{\bm k} = \frac{1}{2} \left( \sqrt{\frac{\omega_\phi}{\omega_\psi}}
 + \sqrt{\frac{\omega_\psi}{\omega_\phi}}  \right) \ ,\qquad
  V_{\bm k} = \frac{1}{2} \left( \sqrt{\frac{\omega_\phi}{\omega_\psi}}
 - \sqrt{\frac{\omega_\psi}{\omega_\phi}}  \right)\ ,
\end{eqnarray}
and time dependent functions
\begin{eqnarray}
f(t) = \frac{\sin \frac{\omega_\phi -\omega_\psi}{2} t}{\frac{\omega_\phi -\omega_\psi}{2}} 
e^{-\frac{i}{2}(\omega_\phi -\omega_\psi)t}\ ,\qquad
g(t) = \frac{\sin \frac{\omega_\phi +\omega_\psi}{2}t}{\frac{\omega_\phi +\omega_\psi}{2}} 
e^{-\frac{i}{2}(\omega_\phi +\omega_\psi)t}\ .
\end{eqnarray}
Note that the relation $U_{\bm k}^2 -V_{\bm k}^2 =1$ holds.

\subsection{Conversion probability for a single photon}

Let us consider an initial state consisting of a single photon with wave vector ${\bm k}$; that is, 
$|{\bm k}_\psi \rangle = b^\dagger_{\bm k}|0 \rangle$.
After the state evolves, it should be projected onto
the axion state $|{\bm k}_\phi \rangle = a^\dagger_{\bm k}|0 \rangle$.
We therefore focus on calculating the transition rate
 \begin{eqnarray}
&&P \left( \gamma \rightarrow \phi \right) 
=\left| \langle {\bm k}_\phi |  \exp[-i \int^t dt H_I (t)] |{\bm k}_\psi \rangle\right|^2  \nonumber\\
&& = \left| \langle 0| a_{\bm k}\,e^{-iQ}\,b_{\bm k}^\dagger |0 \rangle  \right|^2 \nonumber\\
&& = \left| \frac{1}{2}gB U_{\bm k} f^*(t) -\frac{1}{3!}\left( \frac{1}{2}gB \right)^3 U_{\bm k} f^*(t)\Bigl\{ U_{\bm k}^2 |f(t)|^2 + 10 V_{\bm k}^2 |g(t)|^2 \Bigr\} \right. \nonumber\\
&&\left. \qquad + \frac{1}{5!} \left( \frac{1}{2}gB \right)^5 U_{\bm k} f^*(t) \Bigl\{U_{\bm k}^4 |f(t)|^4 + 62 U_{\bm k}^2 V_{\bm k}^2 |f(t)|^2 |g(t)|^2 + 197 V_{\bm k}^4 |g(t)|^4 \Bigr\} - \cdots \right|^2 \ .\quad \qquad 
 \end{eqnarray}
Thus, the leading-order conversion probability is given by
\begin{eqnarray}
    P \left( \gamma \rightarrow \phi \right) 
  =  \left|  \frac{1}{2}gB U_{\bm k} f^*(t) \right|^2
  \simeq  \left(\frac{1}{2} gB L \right)^2 \frac{\sin^2 \frac{m^2}{4k }L}{\left(\frac{m^2}{4 k } L \right)^2}\,.
\end{eqnarray}
The remaining terms represent corrections to this leading-order result.
Similarly, we obtain
\begin{eqnarray}
&&  P \left( \gamma \rightarrow \gamma \right) = \left| \langle{\bm k}_\psi |  \exp[-i \int^t dt H_I (t)] |{\bm k}_\psi \rangle\right|^2 \nonumber \\
&& = \left| 1 - \frac{1}{2!}\left( \frac{1}{2}gB \right)^2 \left\{ U_{\bm k}^2 |f(t)|^2 + 3V_{\bm k}^2 |g(t)|^2 \right\} 
\right. \nonumber\\
&&\left. \qquad 
+ \frac{1}{4!} \left( \frac{1}{2}gB \right)^4 \left\{ 
U_{\bm k}^4 |f(t)|^4 + 25 U_{\bm k}^2 V_{\bm k}^2 |f(t)|^2 |g(t)|^2 + 33 V_{\bm k}^4 |g(t)|^4 \right\} + \cdots \right|^2 \nonumber\\
&&\simeq 1\,.
 \end{eqnarray}

In this section, we have calculated the conversion probability using a quantum field-theoretical approach and reproduced the conventional result at leading order. We are now in a position to take into account the nonclassical photons in the conversion process.

\section{Photon-axion conversion in squeezed coherent states}

In this section, we consider a coherent state and photons-added squeezed coherent state.  We see that the resultant conversion probability is significantly enhanced.

\subsection{Coherent states and squeezed coherent states}
First, we consider the coherent state of photons. 
Experimentally, coherent states are generated by the classical currents ${\bf j}({\bm x})$. By usign Eqs.~(\ref{psi-A}) and (\ref{operator_psi}), we find
\begin{eqnarray}
-i\int dt\int d^3 x\,{\bf j} ({\bm x})\cdot {\bf A}({\bm x})
&=&\frac{-i}{(2\pi)^{3/2}}\int dt\int d^3 x\,j_{\parallel}({\bm x})
\int d^3 k\left[ \psi_{\bm k}(t) b_{\bm k} \, 
+ \psi_{\bm k}^* (t) b_{-\bm k}^\dagger \right]e^{i\bm{k}\cdot\bm{x}}\nonumber\\
&\equiv&\int d^3 k\left(\beta_{\bm k} b_{\bm k}^{\dagger} 
    - \beta^*_{\bm k} b_{-\bm k} \right)
\end{eqnarray}
where $j_{\parallel}$ is the component of the current parallel to the magnetic field. And 
in the last equality, we performed a change of variables ${\bm k}\rightarrow {-\bm k}$. Then we find
\begin{eqnarray}
\beta_{\bm k}=\frac{-i}{(2\pi)^{3/2}}\int dt\int d^3 x\,j_{||}({\bm x})
\,\psi_{\bm k}^* (t)  e^{-i\bm{k}\cdot\bm{x}}\,.
\end{eqnarray}
Hereafter, we focus on a single mode ${\bm k}$. By using a unitary operator called the displacement operator:
\begin{eqnarray}
    D(\beta_{\bm k})  = \exp\left[ \beta_{\bm k} b_{\bm k}^{\dagger} 
    - \beta^*_{\bm k} b_{\bm k} \right]\,,
\end{eqnarray}
the coherent state $|\beta_{\bm k} \rangle$ is constructed as
\begin{eqnarray}
    |\beta_{\bm k} \rangle = D(\beta_{\bm k}) |0 \rangle  \ . 
\end{eqnarray}
This is an eigenstate of the annihilation operator $b_{\bm k}$, satisfying
\begin{eqnarray}
    b_{\bm k}  |\beta_{\bm k} \rangle  
   = \beta_{\bm k}  |\beta_{\bm k} \rangle  \ . 
\end{eqnarray}
Note that two different coherent states 
are not orthogonal to each other; in fact
\begin{eqnarray}
 |\langle  \alpha_{\bm k} |\beta_{\bm k} \rangle|^2  
 = e^{-|\alpha_{\bm k} -\beta_{\bm k}|^2} \ .
\end{eqnarray}

%It is known that the nonlinearity in electrodynamics can produce the squeezed photons.
The squeezed state can be constructed by the squeezing operator
\begin{eqnarray}
    S(\zeta_{\bm k}) = \exp\left[ -\frac{1}{2} \zeta^*_{\bm k} b_{\bm k}^2 +\frac{1}{2} \zeta_{\bm k} b_{\bm k}^{\dagger 2}\right] \ ,\qquad  \zeta_{\bm k} = r_{\bm k} e^{i\varphi_{\bm k}}
\end{eqnarray}
as
\begin{eqnarray}
    |\zeta_{\bm k} \rangle = S(\zeta_{\bm k}) |0 \rangle  \ . 
\end{eqnarray}
It is useful to note that 
\begin{eqnarray}
   b_{\bm k} (\zeta_{\bm k} ) = S^{\dagger}(\zeta_{\bm k}) b_{\bm k}  S(\zeta_{\bm k})
   = b_{\bm k}  \cosh r_{\bm k} + b_{\bm k} ^\dagger e^{i\varphi_{\bm k}} \sinh r_{\bm k}\ . 
\end{eqnarray}
The squeezed coherent state is constructed as $|\zeta_{\bm k} , \beta_{\bm k} \rangle =S(\zeta_{\bm k}) D(\beta_{\bm k})|0\rangle $.
We can also consider a single photon added to the squeezed coherent state $b_{\bm k} ^\dagger |\zeta_{\bm k} ,\beta_{\bm k} \rangle$, and  states with more photons added~\cite{Agarwal:1991zz,Zavatta:2004xxb,Fadrny:2024fef,Thapliyal:2017hms}.

\subsection{Photon-axion conversion probability}

The coherent state is the quantum state that is closest to a classical state. 
Even in the case of a coherent state, we find an enhancement in the conversion probability. 
Let us assume that the initial state of photons is a coherent state $|\beta \rangle $ .
After evolving the state, we evaluate the overlap with
a state consising of an axion added to a photon coherent state, $|{\bm k}_\phi \rangle = a^\dagger_{\bm k}|\alpha \rangle$.
At leading order, we have
 \begin{eqnarray}
 \left| \langle  \alpha , {\bm k}_\phi |  \exp[-i \int^t dt H_I (t)] |\beta \rangle\right|^2   &&= \left| \langle \alpha| a_{\bm k} e^{-iQ} |\beta \rangle  \right|^2 \nonumber\\
&& \simeq  \left(\frac{1}{2} gB \right)^2 U_{\bm k}^2
\left| f^*(t) \right|^2  \left|\beta  \right|^2 \ e^{-|\alpha -\beta|^2}
 \end{eqnarray}
The probability becomes maximum for $\alpha =\beta$.
Notice that we have an enhancement factor of $|\beta|^2$, which corresponds to the laser intensity.

We can consider various non-classical states. Among them, the squeezed state is relatively easy to realize experimentally. 
What we would like to consider is a single photon added to the squeezed coherent state for photons denoted as
$|{\bm k}_\psi ,\zeta ,\beta \rangle=A b_k^\dagger|\zeta ,\beta \rangle $. 
After evolving the state, we evaluate the overlap with a single axion added to the squeezed coherent state $|{\bm k}_\phi , \zeta ,\beta\rangle = a^\dagger_{\bm k}|\zeta ,\beta \rangle$.
Here , the normalization constant $A$ is given by
\begin{align}
    A=\left[
    \cosh^2 r+|\beta|^2\left(
    \cosh 2r + \cos\left[2\arg\beta-\varphi\right] \, \sinh 2r
    \right)
    \right]^{-1/2} \ .
\end{align}
In this case, we can deduce the following leading order result
 \begin{eqnarray}
&&A^2 \left| \langle {\bm k}_\phi ,\zeta,\beta  |  \exp[-i \int^t dt H_I (t)] |{\bm k}_\psi ,\zeta ,\beta \rangle\right|^2   = A^2\left| \langle \zeta ,\beta | a_{\bm k} e^{-iQ} b_k^\dagger|\zeta ,\beta \rangle  \right|^2
\nonumber\\
&& \qquad\qquad = A^2\left| \langle \beta|S^\dagger (\zeta)  a_{\bm k} S(\zeta ) S^\dagger (\zeta) e^{-iQ}S(\zeta ) S^\dagger (\zeta) b_k^\dagger S(\zeta )|\beta \rangle  \right|^2
\nonumber\\
&& \qquad\qquad =  \left(\frac{1}{2} gB  \right)^2 
 U_{\bm k}^2 \left|f^* \right|^2 
 \left[ \cosh^2 r +|\beta|^2\left( \cosh 2r + \sinh 2r \cos (2\delta -\varphi )  \right)\right]  \ ,
 \end{eqnarray}
where we defined $\beta =|\beta| e^{i\delta}$ and $\delta$ is a phase. 
Hence, we have an enhancement by a factor $|\beta|^2 e^{2 r}$.
We can interpret this as the conversion process occuring in an environment of squeezed coherent photons.
Conventionally, the squeezing of a photon is often measured in decibels (dB), defined as $10\log_{10} e^{2r}$.
Currently, it is possible to produce 8 dB or more of photon squeezing in the labotatory~\cite{Kashiwazaki:2023pxy}. 
It should be emphasized that the conversion of axions into photons is enhanced by squeezing.

%Of course, this is not the end of the story. We can generalize the quantum state in various ways. 
%For example, we can consider the cat state
%\begin{eqnarray}
 %   |\psi \rangle  
%    = |\alpha \rangle + e^{i\Phi}|\beta\rangle \ ,
%\end{eqnarray}
%where $\Phi$ is a phase. 
%We can expect various interference patterns for the conversion of photons into axions. 
%We will explore a detailed analysis in a series of papers.
%It is also worth proposing a concrete experimental realization of our findings. 

\section{Conclusion}

In order to detect ultra-light axions, which may account for dark matter and/or dark energy, we expect quantum sensing to play an important role.
In this paper, we focused on the conversion phenomenon.
In the presence of a magnetic field, a photon can be converted into an axion, and vice versa. In order to enhance the sensitivity of detection methods utilizing the conversion phenomenon, 
non-trivial quantum state of photons can be useful as is well known in quantum sensing. However, since the conversion is conventionally treated as classical wave dynamics,
it is not apparent how to incorporate quantum nature such as entanglement. In this work, we formulated photon-axion conversion
in a quantum field theoretical framework. As a result, we successfully derived the conversion probability for a  squeezed coherent state of photons. 
As a consequence, it turns out that the photon-axion conversion probability is enhanced for a  squeezed coherent state.

Note that axions might also be in a squeezed state~\cite{Kanno:2021vwu}. Hence, further enhancement may be expected in the case of axion dark matter cases~\cite{Masaki:2019ggg}.
It is straightforward to incorporate plasma and one-loop effects into our analysis. 
Our approach can also be extended to photon-graviton conversion~\cite{Ikeda:2025uae}.
This may be closely related to previous works in a de Sitter universe~\cite{Kanno:2022ykw,Kanno:2022kve,Kanno:2023fml}.
It might be useful for the detection of high-frequency gravitational waves, where various conversion processes are used~\cite{Ito:2019wcb,Ejlli:2019bqj,Ito:2020wxi,Berlin:2021txa,Domcke:2022rgu,Ito:2022rxn}. 
We will report these calculaions in a series of separate papers.

\section*{Acknowledgments}
We would like to thank Youka Kaku for useful comments.
 S.~K. was supported by the Japan Society for the Promotion of Science (JSPS) KAKENHI Grant Numbers JP22K03621, JP22H01220, 24K21548 and MEXT KAKENHI Grant-in-Aid for Transformative
Research Areas A “Extreme Universe” No. 24H00967.
J.\ S. was in part supported by JSPS KAKENHI Grant Numbers  JP23K22491, JP24K21548, JP25H02186.

%\printbibliography 
\bibliography{axion} 
\bibliographystyle{unsrt}
\end{document}